\documentclass[hyper]{JHEP} 

\usepackage{epsfig}

\newcommand\fverb{\setbox\pippobox=\hbox\bgroup\verb}

\newcommand\fverbdo{\egroup\medskip\noindent%

            \fbox{\unhbox\pippobox}\ }

\newcommand\fverbit{\egroup\item[\fbox{\unhbox\pippobox}]}

\newbox\pippobox

\title{Carrollian Limit of Bimetric Gravity}
\author{J. Kluso\v{n}\\
Department of
Theoretical Physics and Astrophysics\\
Faculty of Science, Masaryk University\\
Kotl\'{a}\v{r}sk\'{a} 2, 611 37, Brno\\
Czech Republic\\
E-mail: \email{klu@physics.muni.cz}} \preprint{}

 \abstract{We expand bimetric theory of gravity 
in  small speed of light limit. We find electric and magnetic Carrollian form of the action and discuss their properties.}

\def\mM{\mathcal{M}}

\def\bA{\mathbf{A}}

\def\be{\begin{equation}}

\def\ee{\end{equation}}

\def\bea{\begin{eqnarray}}

\def\eea{\end{eqnarray}}

\def\tr{\mathrm{tr}\, }

\def\tr{\mathrm{Tr}}

\def\mE{\mathcal{E}}

\newcommand{\mK}{\mathcal{K}}

\newcommand{\mF}{\mathcal{F}}

\newcommand{\mG}{\mathcal{G}}

\def \bA{\mathbf{A}}

\newcommand{\mL}{\mathcal{L}}

\begin{document}
\section{Introduction and Summary}
Bimetric theory of gravity describes non-linear interaction 
of massive spin two field with gravity 
\cite{Hassan:2011zd,Hassan:2012wr}. In other words, 
bimetric theories of gravity are based on the idea to join  two
tensors $g_{\mu\nu}$ and $f_{\mu\nu}$ in a symmetric way when
each tensor has its own Einstein-Hilbert action and when these metrics are coupled through a non-derivative mass term. The presence of
this term reduces the separate coordinate invariance to a single
one \cite{Salam:1969rq,Isham:1970gz}.
In bimetric gravity the interaction is mediated by massive and massless spin two modes. For large spin two masses bimetric gravity is in agreement with all local gravity test due to the  
Yukawa suppression of the massive graviton mode in the gravitational potential in Newtonian approximation.
It was also shown in \cite{Comelli:2011zm,Volkov:2011an,vonStrauss:2011mq}
that bimetric theory possesses cosmological solutions with accelerated expansion even if the cosmological constant is zero
\footnote{For recent review, see \cite{Schmidt-May:2015vnx}.}. In other words bimetric or more generally multimetric gravity are very interesting models of gravity with rich theoretical structure
and with potentially phenomenological applications so that it is natural to study them from different
point of view. 

One such  possibility is to analyze how bimetric gravity depend on the speed of light $c$.  
Recently General Relativity was analyzed in the limit $c\rightarrow \infty$ which leads to very important
developments known as Newton-Cartan gravity, for review and extensive list of references, 
see \cite{Hartong:2022lsy,Hansen:2020pqs}. These works were based on  expansion of General Relativity 
in inverse powers of $c$ that lead to the replacement of Lorentz geometry with Newton-Cartan geometry 
at leading order of expansion however this treatment is more general and leads to many interesting results \cite{Hartong:2022lsy,Hansen:2020pqs}.

It is also interesting to study opposite limit of small speed of light $c$. In this limit the Poincare
group contracts to Carroll group when the speed of light goes to zero which has interesting consequences for dynamics and kinematics. In particular, particles with non-zero energy 
cannot move in space-time \cite{deBoer:2021jej} and for these particles there are also no interactions between spatially separated events. In other words Carroll limit is ultra local limit 
\cite{Henneaux:1979vn}. Then we can interpret an expansion in the small $c$ as expansion around singular Carroll point.

In a very nice paper \cite{Hansen:2021fxi} 
systematic expansion of General Relativity in the small speed of light limit was studied with  analogy
with non-relativistic expansion \cite{Hansen:2020pqs}. This procedure is based on definition of  
 pre-ultra local variables (PUL)
\cite{Hansen:2020pqs,Hansen:2019pkl} and this expansion allows systematic study of dynamics of leading order (LO) and next to leading order (NLO) terms in the action but also terms of higher order.

It is important to stress that the analysis introduced in  \cite{Hansen:2020pqs} is very general and  can be used for any covariant theory of gravity. For example, this procedure was systematically used in \cite{Tadros:2023teq} where PUL expansion of quadratic gravity was analyzed. 
In this paper we perform this PUL expansion in case of bimetric gravity
\footnote{This problem was studied recently in \cite{Ekiz:2022wbi} from different point of view.}. Note that bimetric gravity is 
formulated with two metrics $g_{\mu\nu}$ and $f_{\mu\nu}$ where these metrics are coupled by specific potential 
that contains the square root of the matrix
$g^{\mu\nu}f_{\nu\rho}$. This is rather awkward structure which can be avoided when
we formulate bimetric or generally multimetric theories of gravity in vielbein formalism
\cite{Hinterbichler:2012cn} and this is the formulation that will be useful for Carroll expansion. 
In more details, we introduce two set of PUL variables for metric $g_{\mu\nu}$ and $f_{\mu\nu}$ respectively. Since the kinetic terms for  these metrics have standard Einstein-Hilbert form it turns out that the Carroll expansions of these terms is completely the same as in the case of General Relativity. On the other hand using
PUL variables in interaction term we find that it does not depend on $c$ at all which has crucial impact on the Carrollian expansion of bimetric gravity. In fact, performing PUL expansion of bimetric gravity we can identify, following 
\cite{Hansen:2020pqs} electric and magnetic Carrollian limits in terminology introduced in \cite{Henneaux:2021yzg}. We find that electric limit corresponds to two non-interacting parts where each part has the same form as in case of General Relativity while the magnetic limit contains interaction term between two metric $g_{\mu\nu}$ and $f_{\mu\nu}$.  Further,  all other terms of higher order in $c^2$ have the same form as in case of General Relativity
and   the equations of motion of Carroll limit of bimetric gravity have the same forms as in case of General Relativity  so that the discussion performed in \cite{Hansen:2020pqs} holds for each metric separately.  
 
 We can outline results derived in this paper as follows. We found PUL expansion of bimetric theory of gravity in the small $c^2$ expansion. We identified electric Carroll limit that corresponds to 
 non-interacting terms for each metric $g_{\mu\nu}$ and $f_{\mu\nu}$ respectively. The interaction term between these metrics has meaning in case of magnetic Carrollian limit only where however the dynamics is trivial. On the other hand it would be interesting to study solutions of equations of motion of 
 magnetic Carrollian bimetric gravity and compare them with small speed of light limit of solutions of bimetric gravity. We hope to return to this problem in future.

This paper is organized as follows. In the next section (\ref{second}) we introduce basic form of  bimetric
gravity. Then in section (\ref{third}) we introduce PUL parametrization of metrics $g_{\mu\nu}$ and $f_{\mu\nu}$ respectively. We review, following \cite{Hansen:2020pqs}, expansion 
of Christoffell connection and Ricci scalar in terms of PUL variables. Then with the help of these formulas we find  Carroll  expansion of bimetric gravity.

\section{Review of Bimetric Gravity}\label{second}
As the name suggests bimetric gravity is specific model of theory of gravity with two dynamical metrics $g_{\mu\nu}$ and $f_{\mu\nu}$ which are coupled by interaction terms. Thanks to the specific form of interaction between these two metrics it is most convenient to formulate  bimetric gravity with the help of vielbeins
 \cite{Hinterbichler:2012cn}.
In vielbein formulations we consider metric $g_{\mu\nu}$ and 
$f_{\mu\nu}$ written with the help of vielbeins $E_\mu^{ \ A}$ and
$F_{\mu}^{ \ A}$  
\begin{equation}
g_{\mu\nu}=E_\mu^{ \ A}E_\nu^{ \ B}\eta_{AB} \ , \quad
f_{\mu\nu}=F_\mu^{ \ A}F_\nu^{ \ B}\eta_{AB} \ , \quad 
\eta_{AB}=\mathrm{diag}(-1,1,\dots,1) \ , 
\end{equation}
where $\mu,\nu=0,1,\dots,d$ and $A,B=0,1,\dots,d$. Then  the action for bimetric gravity  has the form 
 \cite{Hinterbichler:2012cn}
\begin{eqnarray}
	S&=&\frac{M_g^{2}}{2} \int d^{d+1}x (\det E)R[E] + \frac{M_f^{2}}{2}
	\int d^{d+1}x (\det F) R[F] -
	\nonumber \\
	&-&\mu^2 \int d^{d+1}x \sum_{n=0}^4 \beta_n (\det E)
	S_n(\Theta F) \ ,  \nonumber \\
\end{eqnarray}
where $\mu^2=\frac{1}{8}m^2 M_{g}^{2}$ and where $M_g^2$ and $M_f^2$ will be explicitly 
defined in (\ref{third}) section.  Further,  $S_n$ are symmetric polynomials whose explicit
definitions were found in \cite{Hinterbichler:2012cn}. It was
shown there that they can  be written in terms of traces of the matrix
$M^\mu_{ \ \nu}\equiv \Theta^\mu_{ \ A}F_\nu^{ \ A}$  as
\begin{eqnarray}
	S_0(M)&=&1 \ , \nonumber \\
	S_1(M)&=&[M] \nonumber \\
	S_2(M)&=&\frac{1}{2!}([M]^2-[M^2])
	\ ,
	\nonumber \\
	S_3(M)&=&\frac{1}{3!}
	([M]^3-3[M][M^2]+2[M^3]) \ ,
	\nonumber \\
	S_4(M)&=&
	\frac{1}{4!}
	([M]^4-6[M^2][M]^2+8[M][M^3]
	+3[M^2]^2-6[M^4]) \ ,
	\nonumber \\
\end{eqnarray}
where $[M]$ means the trace of the matrix $M^\mu_{ \ \nu}$. Finally,  following 
\cite{Hansen:2021fxi} we introduced $\Theta^\mu_{ \ A}$ as inverse to $E_\mu^{ \ A}$
\begin{equation}
E_\mu^{ \ A}\Theta^\mu_{ \ B}=\delta^A_B \ . 
\end{equation}
 This specific form of interaction term has crucial impact on consistency of this theory as was argued in 
 \cite{Hassan:2011zd,Hassan:2012wr}.  
  In this short note we would like to study Carrollian limit of this theory. To do this we review basic facts about pre-ultralocal expansion of metric components $g_{\mu\nu}$ and $f_{\mu\nu}$. 

\section{Pre-Ultralocal Expansion and Bimetric Gravity}\label{third}
We start this section with the review of  pre-ultra local  parametrization of vielbein and metric, following \cite{Hansen:2021fxi}. 
For simplicity we discuss the metric $g_{\mu\nu}$ only since
in case of the metric $f_{\mu\nu}$ the situation is identical.

Let us consider Lorentzian metric $g_{\mu\nu}$ and its inverse $g^{\mu\nu}$ written in  PUL parametrization \cite{Hansen:2021fxi}
\begin{equation}\label{PULparam}
	g_{\mu\nu}=-c^2 T_{\mu}T_{\nu}+\Pi_{\mu\nu} \ , \quad 
g^{\mu\nu}=-\frac{1}{c^2}V^\mu V^\nu+\Pi^{\mu\nu} \ ,
\end{equation}	
where we have time-like vector and one form $V^\mu$ and $T_\mu$ and we have implicitly 
introduced speed of light $c$ as expansion parameter. 
Now using  the definition  $g_{\mu\nu}g^{\nu\rho}=\delta_\mu^\rho$
and comparing terms of different order in $c^2$ we obtain following set of conditions that these variables have to obey
\begin{equation}\label{relation1x}
	\Pi_{\mu\nu}V^\nu=0 \ , \quad T_\mu \Pi^{\mu\nu}=0 \ , \quad 
	V_\mu T^\mu=-1 \ , \quad 
	-V_\mu T^\nu+\Pi_{\mu\rho}\Pi^{\rho\nu}=\delta_\mu^\nu \ . 
\end{equation}
It is convenient to formulate PUL expansion of metric with the help of vielbein $E_\mu^{ \ A}$
and its inverse $\Theta^\nu_{ \ A}$. Recall that obey the conditions 
\begin{eqnarray}
	E_\mu^{ \ A}\Theta^\mu_{ \ B}=\delta^A_B \ , \quad 
	E_\mu^{ \ A}\Theta^\nu_{ \ A}=\delta_\mu^\nu \ . 
\nonumber \\
\end{eqnarray}
In PUL parametrization  (\ref{PULparam}) we find that $E_\mu^{ \ A}$ and $\Theta^\mu_{ \ B}$ are equal to
\begin{equation}
	E_\mu^{ \ A}=(cT_\mu,E_\mu^{ \ a}) \ , \quad 
	\Theta^\mu_{ \ A}=(-\frac{1}{c}V^\mu,\Theta^\mu_{ \ a}) \ .
\end{equation}
so that
\begin{equation}
	\Pi_{\mu\nu}=E_\mu^{\ a}E_\nu^{ \ b}\delta_{ab} \ ,  \quad 
\Pi^{\mu\nu}=\Theta^\mu_{ \ a}\Theta^\nu_{ \ b}\delta^{ab} \ , 
\end{equation}
where $A,B=0,1,\dots,D$ are spacetime frame indices and $a,b=1,\dots,d$ are space frame indices. 

Following again \cite{Hansen:2021fxi} we   presume that PUL variables can be analytically expanded in $ c^2$ as
\begin{eqnarray}
&&	V^\mu=v^\mu+c^2 M^\mu+\mathcal{O}(c^4) \ , \nonumber \\
&&	\Theta^\mu_{ \ a}=\theta^\mu_{ \ a}+c^2\pi^\mu_{ \ a}+\mathcal{O}(c^4) \ , \nonumber \\
&&T_\mu=\tau_\mu+\mathcal{O}(c^2) \ , \nonumber \\
&&E_\mu^{ \ a}=e_\mu^{ \ a}+\mathcal{O}(c^2) \  \nonumber \\	
\end{eqnarray}
so that
\begin{eqnarray}
&&\Pi_{\mu\nu}=E_\mu^{ \ a}E_\nu^{ \ b}\delta_{ab}=
h_{\mu\nu}+\mathcal{O}(c^2) \ , \quad h_{\mu\nu}=e_\mu^{ \ a}e_\nu^{ \ b}\delta_{ab}  \ , \nonumber \\
&&\Pi^{\mu\nu}=h^{\mu\nu}+c^2\Phi^{\mu\nu}+\mathcal{O}(c^4) \ , 
\quad \Phi^{\mu\nu}=e^\mu_{ \ a}\pi^\nu_{ \ b}\delta^{ab}+
\pi^{\mu}_{ \ a}e^\nu_{ \ b}\delta^{ab} \ . \nonumber \\
\end{eqnarray}
It is important to stress that we have not included subleading terms in the expansion of
$T_\mu$ and $E_\mu^{ \ a}$ due to the fact that these variables are determined
by variables in expansion of $V^\mu$ and $E^\mu_{ \ a}$ using the relations (\ref{relation1x})
and we obtain 
\begin{equation}
	\tau_\mu v^\mu=-1 \ , \quad \tau_\mu h^{\mu\nu}=0 \ , \quad 
	h_{\mu\nu}v^\nu=0 \ , \quad -\tau_\mu v^\nu+h_{\mu\rho}h^{\rho\nu}=\delta_\mu^
	\nu \ . 
\end{equation}
These variables define Carroll geometry for metric $g_{\mu\nu}$. In the same way we proceed with 
$f_{\mu\nu}$ when we write
\begin{equation}
	f_{\mu\nu}=-c^2 U_{\mu}U_{\nu}+\Sigma_{\mu\nu} \ , \quad 
	f^{\mu\nu}=-\frac{1}{c^2}W^\mu W^\nu+\Sigma^{\mu\nu} \ , 
\end{equation}	
where we have time-like vector and one form $W^\mu$ and $U_\mu$.
We again have 
\begin{equation}\label{relation1}
	\Sigma_{\mu\nu}W^\nu=0 \ , \quad U_\mu \Sigma^{\mu\nu}=0
\ ,  \quad 	U_\mu W^\mu=-1 \ , \quad  
	-U_\mu W^\nu+\Sigma_{\mu\rho}\Sigma^{\rho\nu}=\delta_\mu^\nu \ . 
\end{equation}
As in case metric $g_{\mu\nu}$ we  introduce
vielbein $F_\mu^{ \ A}$ and its inverse $\Psi^\nu_{ \ A}$
so that
\begin{eqnarray}
	f_{\mu\nu}=F_\mu^{ \ A}F_\nu^{ \ B}\eta_{AB} \ , 
	\quad f^{\mu\nu}=\Psi^\mu_{ \ A}\Psi^\nu_{ \ B}\eta^{AB}
\ . 	\nonumber \\
\end{eqnarray}
In terms of PUL parametrization they have the form 
\begin{equation}
	F_\mu^{ \ A}=(cU_\mu,F_\mu^{ \ a}) \ , \quad 
	\Psi^\mu_{ \ A}=(-\frac{1}{c}W^\mu,\Psi^\mu_{ \ a}) \ 
\end{equation}
so that
\begin{equation}
	\Sigma_{\mu\nu}=F_\mu^{\ a}F_\nu^{ \ b}\delta_{ab} \ , \quad 
	\Sigma^{\mu\nu}=\Psi^\mu_{ \ a}\Psi^\nu_{ \ b}\delta^{ab} \ .
\end{equation}
We further presume that  PUL vielbein $F_\mu^{ \ A}$ can be analytically expanded in $c^2$
\begin{eqnarray}
&&	W^\mu=w^\mu+c^2 N^\mu+\mathcal{O}(c^4) \ , \nonumber \\
&&	\Psi^\mu_{ \ a}=\psi^\mu_{ \ a}+c^2\rho^\mu_{ \ a}+\mathcal{O}(c^4) \ , \nonumber \\
&&	U_\mu=u_\mu+\mathcal{O}(c^2) \ , \nonumber \\
&&	F_\mu^{ \ a}=f_\mu^{ \ a}+\mathcal{O}(c^2) \ . \nonumber \\	
\end{eqnarray}
where the  leading order variables  satisfy 
\begin{equation}
	u_\mu w^\mu=-1 \ , \quad u_\mu k^{\mu\nu}=0 \ , \quad 
	k_{\mu\nu}w^\nu=0 \ , \quad -u_\mu w^\nu+k_{\mu\rho}k^{\rho\nu}=\delta_\mu^
	\nu \ . 
\end{equation}
Now we can proceed to the interaction term between two vielbeins
$\Theta^\mu_{ \ A}$ and $F_\mu^{ \ B}$. In fact, this interaction terms is
function of the matrix $M_\mu^{ \ \nu}$ only where $M_\mu^{ \ \nu}$ is
defined as
\begin{equation}
	M^\mu_{ \ \nu}=\Theta^\mu_{ \ A}F_\nu^{ \ A}=
	-V^\mu U_\nu+\Theta^\mu_{ \ a}F_\nu^{ \ a} \ 
		\end{equation}
and we find crucial result that this matrix does not depend on 
negative powers of $c^2$. Note that in the limit $c\rightarrow 0$ this matrix is equal to
\begin{equation}\label{defmmunu}
m^\mu_{ \ \nu}=\lim_{c\rightarrow 0}M^\mu_{ \ \nu}=
 -v^\mu u_\nu+\theta^\mu_{ \ a}f_\nu^{ \ a} \ . 
\end{equation} 
Before we proceed to the Carroll expansion of bimetric gravity we should determine 
value of $\det g$ that in PUL variables it is equal to  
\begin{equation}
	\det g=\det (-c^2T_\mu T_\nu+\Pi_{\mu\nu}) \ . 
\end{equation}
In order to calculate this determinant we rewrite it into an equivalent form 
\begin{eqnarray}
&&	\det g
=\det (\bA_{\mu\nu}+(1-c^2)T_\mu T_\nu)=
\det \bA \det (\delta^\mu_\rho+\bA^{\mu\sigma}T_\sigma T_\rho(1-c^2))=\nonumber \\
&&=\det \bA \det (\delta^\mu_\rho+V^\mu T_\rho(1-c^2)) \ ,  \nonumber \\
\end{eqnarray}
where we introduced the matrix $\bA_{\mu\nu}$ 
\begin{equation}
\bA_{\mu\nu}=-T_\mu T_\nu+\Pi_{\mu\nu}
\end{equation}
and its inverse 
\begin{equation}
\bA^{\mu\nu}=-V^\mu V^\nu+\Pi^{\mu\nu} \ , \quad \bA_{\mu\nu}\bA^{\nu\rho}=\delta_\mu^\nu
\end{equation}
and we also used the fact that 
\begin{equation}
	\bA^{\mu\nu}T_\nu=V^\mu \ .
\end{equation}
Then using the relation $\det X=\exp \tr \ln X$ that holds for any matrix we easily find that 
\begin{eqnarray}
	\det (\delta^\mu_\rho+V^\mu T_\rho(1-c^2))=c^2
\end{eqnarray}
so that we obtain final result 
\begin{equation}
	\det g=c^2 \det (-T_\mu T_\nu+\Pi_{\mu\nu})\equiv c^2 \mathcal{E}^2 \ 
\end{equation}
that will be crucial for Carroll expansion of bimetric gravity. At the same way we have
\begin{equation}
	\det f=c^2\det (-U_\mu U_\nu+\Sigma_{\mu\nu})\equiv c^2\mF^2 \ . 
\end{equation}

\subsection{Carroll Compatible Connection}
Following \cite{Hansen:2021fxi} we introduce Carroll compatible connection by condition 
\begin{equation}\label{condC}
\stackrel{(C)}{\nabla}_\mu V^\nu=0 \ , \quad \stackrel{(C)}{\nabla}_\rho\Pi_{\mu\nu}=0 \ ,
\end{equation}
where the covariant  derivatives $\stackrel{(C)}{\nabla}$ are defined  with the connection $C_{\mu\nu}^\rho$. These conditions do not fully 
determine connections $C_{\mu\nu}^\rho$. In fact, it can be shown that it has the form 
\cite{Hansen:2021fxi}
\begin{eqnarray}
&&	C^\rho_{\mu\nu}=
	-V^\rho (\partial_\mu T_\nu+\partial_\nu T_\mu)-
V^\rho (T_\mu \mL_V T_\nu+T_\nu \mL_VT_\mu)+
\nonumber \\
&&+\frac{1}{2}\Pi^{\rho\lambda}
[\partial_\mu \Pi_{\nu\lambda}+\partial_\nu \Pi_{\lambda\mu}
-\partial_\lambda \Pi_{\mu\nu}]-
\Pi^{\rho\lambda}T_\nu \mK_{\mu\lambda} \ , \nonumber \\
\end{eqnarray}
where
\begin{equation}
\mK_{\mu\nu}=-\frac{1}{2}\mL_V \Pi_{\mu\nu}	
	\end{equation}
	and where
\begin{eqnarray}
&&	\mL_V T_\nu=
	V^\mu\partial_\mu T_\nu+\partial_\nu V^\mu T_\mu  \ , \nonumber \\
&&\mL_V \Pi_{\mu\nu}=
V^\rho\partial_\rho \Pi_{\mu\nu}+\partial_\mu V^\rho
\Pi_{\rho\nu}+\partial_\nu V^\rho \Pi_{\rho\mu} \ . \nonumber \\
\end{eqnarray}
Then it can be easily shown that 
\begin{eqnarray}
	V^\mu \mK_{\mu\nu}=0
\end{eqnarray}	
and hence $\mK_{\mu\nu}$ is pure spatial tensor.
In order to obtain Carroll gravity as limiting case of General Relativity we have to find relation between $C^\rho_{\mu\nu}$ and the standard Christoffell connection  $\Gamma^\rho_{\mu\nu}$ that
is defined as
\begin{equation}
	\Gamma^\rho_{\mu\nu}=
\frac{1}{2}g^{\rho\sigma}(\partial_\mu g_{\sigma \nu}+
\partial_\nu g_{\sigma \mu}-\partial_\sigma g_{\mu\nu}) \ . 
\end{equation}
Using PUL parametrization of the metric it can be shown that $\Gamma^\rho_{\mu\nu}$ is equal to 
\cite{Hansen:2021fxi}
\begin{equation}
\Gamma^{\rho}_{\mu\nu}=
\frac{1}{c^2}\stackrel{(-2)}{C^\rho_{\mu\nu}}+C^\rho_{\mu\nu}+S^\rho_{\mu\nu}+c^2\stackrel{(2)}{C^\rho_{\mu\nu}} \ , 
\end{equation}
where
\begin{eqnarray}
&&	\stackrel{(-2)}{C^\rho_{\mu\nu}}=-V^\rho \mK_{\mu\nu} \ , 
	S^\rho_{\mu\nu}=\Pi^{\rho\lambda}T_\nu\mK_{\mu\lambda} \ , 
	\nonumber \\	
&&		\stackrel{(2)}{C^\rho_{\mu\nu}}=-\Pi^{\rho\sigma}
	(\partial_\mu T_\sigma T_\nu+\partial_\nu T_\sigma T_\mu-
\partial_\sigma(T_\mu T_\nu)) \ . \nonumber \\
\end{eqnarray}
In the similar way we   can proceed with the Ricci scalar $R_g=g^{\mu\nu}R_{\mu\nu}$ where  Ricci tensor is defined as
\begin{equation}
	R_{\mu\nu}=-\partial_\mu \Gamma^\rho_{\rho\nu}+
	\partial_\rho \Gamma^\rho_{\mu\nu}-
	\Gamma^\rho_{\mu\nu}\Gamma^\lambda_{\rho\nu}+
	\Gamma^\rho_{\rho\lambda}\Gamma^\lambda_{\mu\nu}
\end{equation}
Using PUL parametrization of metric and after some length calculations we find that it is equal to 
\begin{eqnarray}
&&	R_g=\frac{1}{c^2}[\mK^2-\Pi^{\lambda\nu}
\mK_{\mu\lambda}\Pi^{\mu\alpha}\mK_{\nu\alpha}-
2\stackrel{(C)}{\nabla_\nu}(V^\nu\mK)]+ \nonumber \\
&&+[-\Pi^{\lambda \nu}\stackrel{(C)}{R}_{\lambda\nu}+
\Pi^{\lambda\nu}\stackrel{(C)}{\nabla}_\mu(\Pi^{\mu\alpha}T_\lambda \mK_{\nu\alpha})-\Pi^{\lambda\nu}\stackrel{(C)}{\nabla}_\nu(T_\lambda \mK)+\nonumber \\
&&+V^\lambda V^\nu \stackrel{(C)}{\nabla}_\mu(T_{(\nu}\Pi^{\mu\alpha}B_{\lambda)\alpha})-
V^\lambda V^\nu \stackrel{(C)}{\nabla}_\nu(T_{(\mu}\Pi^{\mu\alpha}B_{\lambda)\alpha})]+\nonumber \\
&&+c^2[-\Pi^{\lambda\nu}\stackrel{(C)}{\nabla}_\mu(T_{(\nu}\Pi^{\mu\alpha}B_{\lambda)\alpha})+\Pi^{\lambda\nu}
\stackrel{(C)}{\nabla}_\nu(T_{(\mu}\Pi^{\mu\alpha}B_{\lambda)\alpha})-
\frac{1}{4}B_{\mu\nu}B^{\mu\nu}] \ , 
\nonumber \\
\end{eqnarray}
where
\begin{equation}
	\mK=\Pi^{\mu\nu}\mK_{\mu\nu} \ , \quad 	B^g_{\mu\nu}=\partial_\mu T_\nu-\partial_\nu T_\mu \ . 
\end{equation}
and where
\begin{equation}
	\stackrel{(C)}{R_{\mu\nu}}=
	\partial_\mu C_{\nu\lambda}^\mu-\partial_\nu C^\mu_{\mu\lambda}
	+C^\mu_{\mu\sigma}C_{\nu\lambda}^\sigma-C^\mu_{\nu\sigma}C^
	\sigma_{\mu\lambda} \ . 
\end{equation}
Now using the relation 
\begin{equation}
\stackrel{(C)}{\nabla}_\rho \Pi^{\mu\nu}=
	-V^{(\mu}\Pi^{\nu)\sigma}B_{\sigma\lambda}
	[\delta^\lambda_\rho-V^\lambda T_\rho]
\end{equation}
we obtain 
\begin{eqnarray}
	\Pi^{\lambda \nu}\stackrel{(C)}{\nabla}_\mu(T_\lambda \mK^\mu_\nu)=0 \ , 
	\quad \Pi^{\mu\nu}\stackrel{(C)}{\nabla}_\nu(T_\lambda \mK)=0 \ 
\end{eqnarray}
that allows us to simplify $R_g$ into final form
(up to boundary terms)
\begin{eqnarray}\label{Rg}
	R_g=\frac{1}{c^2}[\Pi^{\lambda\nu}
	\mK_{\mu\lambda}\Pi^{\mu\alpha}\mK_{\nu\alpha}-\mK^2]
	+\Pi^{\lambda \nu}\stackrel{(C)}{R}_{\lambda\nu}+
	c^2
	\frac{1}{4}B^g_{\mu\nu}B^{\mu\nu}_g\ .
	\nonumber \\
\end{eqnarray}
In the same way $R_f$ is equal to 
\begin{equation}\label{Rf}
	R_f=
	\frac{1}{c^2}
	[\Sigma^{\lambda\nu}
	\mM_{\mu\lambda}\Sigma^{\mu\alpha}
	\mM_{\nu\alpha}-\mM^2]+
	\Sigma^{\lambda\nu}\stackrel{(D)}{R}_{\lambda\nu}+c^2
	\frac{1}{4}B^f_{\mu\nu}B^{\mu\nu}_f \ , 
\end{equation}
where
\begin{equation}
	\mM_{\mu\nu}=-\frac{1}{2}\mL_U \Sigma_{\mu\nu} \ , 
	\quad B^f_{\mu\nu}=\partial_\mu U_\nu-\partial_\nu U_\mu \ ,  
\end{equation}
and where $\stackrel{(D)}{R}_{\mu\nu}$ is defined as
\begin{equation}
R^f_{\mu\nu}=\partial_\mu D_{\nu\lambda}^\mu-\partial_\nu D^\mu_{\mu\lambda}
	+D^\mu_{\mu\sigma}D_{\nu\lambda}^\sigma-D^\mu_{\nu\sigma}D^
	\sigma_{\mu\lambda} \ .
\end{equation}
Note that $D^\rho_{\mu\nu}$ defines covariant derivative compatible with constraints
\begin{equation}
	\stackrel{(D)}{\nabla}_{\mu}\Sigma_{\nu\rho}=0 \ , 
	\quad 
	\stackrel{(D)}{\nabla}_\mu W^\nu=0 \ . 
\end{equation}
Now we are ready to proceed to the Carroll expansion of bimetric gravity. 
Recall that the action has the form 
\begin{eqnarray}\label{bi2}
	S&=&\frac{M_g^{2}}{2} \int d^{d+1}x (\sqrt{-g}R_g + \frac{M_f^{2}}{2}
	\int d^{d+1}x \sqrt{-f} R_f-
	\nonumber \\
	&-&\mu^2 \int d^{d+1}x \sum_{n=0}^4 \beta_n \sqrt{-g}
	S_n(M^\mu_{ \ \nu}) \ ,  \nonumber \\
\end{eqnarray}
where 
\begin{equation}
	\mu^2=\frac{1}{8}m^2 M_g^2 \ , \quad
	M_g^2=\frac{c^3}{8\pi G_g} \ , \quad  M_f^2=\frac{c^3}{8\pi G_f} \  ,
\end{equation}
where we included one factor of $c$ into definition of $\sqrt{-\det g}$ and $\sqrt{-f}$ which is equal to
\begin{equation}
	\sqrt{-\det g}=c \mE \ , \quad \sqrt{-\det f}=c \mF \ , 
\end{equation}
where
\begin{equation}
	\mE=\det (-T_\mu T_\nu+\Pi_{\mu\nu}) \ , 
	\quad \mF=\det (-W_\mu W_\nu+\Sigma_{\mu\nu}) \ . 
\end{equation}
Inserting (\ref{Rg}) and (\ref{Rf}) into (\ref{bi2}) we obtain
\begin{eqnarray}
&&	S=\frac{c^2}{16\pi G_g} \int d^{d+1}x \mE 
[\Pi^{\lambda\nu}
\mK_{\mu\lambda}\Pi^{\mu\alpha}\mK_{\nu\alpha}-\mK^2
+c^2\Pi^{\lambda \nu}\stackrel{(C)}{R}_{\lambda\nu}+
c^4
\frac{1}{4}B^g_{\mu\nu}B^{\mu\nu}_g]+\nonumber \\
&&	 + \frac{c^2}{16\pi G_f}
	\int d^{d+1}x \mF 
[\Sigma^{\lambda\nu}
\mM_{\mu\lambda}\Sigma^{\mu\alpha}
\mM_{\nu\alpha}-\mM^2]+
c^2\Sigma^{\lambda\nu}\stackrel{(D)}{R}_{\lambda\nu}+c^4
\frac{1}{4}B^f_{\mu\nu}B^{\mu\nu}_f]	
 -
	\nonumber \\
&&	-\frac{c^4}{64\pi G} \int d^{d+1}x \sum_{n=0}^4 \beta_n \mE
	S_n(M) \equiv 	
		  c^2 \stackrel{(2)}{S}_{LO}+c^4\stackrel{(4)}{S}_{NLO} \ .
	  \nonumber \\
	\end{eqnarray}
We see that the leading order Lagrangian of  bimetric gravity is equal to 
\begin{eqnarray}
	\mL_{LO}=
\frac{1}{16\pi G_g}  \mE 
[\Pi^{\lambda\nu}
\mK_{\mu\lambda}\Pi^{\mu\alpha}\mK_{\nu\alpha}-\mK^2]+
\frac{1}{16\pi G_f}\mF 
[\Sigma^{\lambda\nu}
\mK_{\mu\lambda}\Sigma^{\mu\alpha}
\mM_{\nu\alpha}-\mM^2]
\end{eqnarray}
which is the sum two non-interacting parts. Note that in the limit 
$c\rightarrow 0$ the Lagrangian leads to electric Carroll limit  
\begin{eqnarray}
	\mL_{el}=\lim_{c\rightarrow 0}\mL_{LO}=
	\frac{1}{16\pi G_g}e [h^{\alpha\nu}K_{\mu\lambda}
	h^{\mu\alpha}K_{\nu\alpha}-K^2
]+\frac{1}{16\pi G_f}f [k^{\mu\nu}M_{\mu\lambda}
k^{\mu\alpha}M_{\nu\alpha}-M^2] \ . \nonumber \\
\end{eqnarray}

 More interesting situation occurs
in case of the magnetic limit that is defined in the following way. 
Let us rewrite the PUL form of bimetric gravity action into the form 
\begin{eqnarray}\label{PULmetric}
&&	S=\frac{c^4}{16\pi G_g} \int d^{d+1}x \mE 
	[\frac{1}{c^2}\mK_{\mu\nu}\mG^{\mu\nu\rho\sigma}\mK_{\rho\sigma}
	+\Pi^{\lambda \nu}\stackrel{(C)}{R}_{\lambda\nu}+
	+c^2
	\frac{1}{4}B^g_{\mu\nu}B^{\mu\nu}_g]+\nonumber \\
&&	+ \frac{c^4}{16\pi G_f}
	\int d^{d+1}x \mF 
	[\frac{1}{c^2}\mM_{\mu\nu}\tilde{\mG}^{\mu\nu\rho\sigma}\mM_{\rho\sigma}
+	\Sigma^{\lambda\nu}\stackrel{(D)}{R}_{\lambda\nu}+c^2
	\frac{1}{4}B^f_{\mu\nu}B^{\mu\nu}_f]	
	-
	\nonumber \\
&&	-\frac{c^4}{64\pi G} \int d^{d+1}x \sum_{n=0}^4 \beta_n \mE
	S_n(M) \ ,  \nonumber \\
\end{eqnarray}
where
\begin{eqnarray}
	\mG^{\mu\nu\rho\sigma}=
	\frac{1}{2}(\Pi^{\mu\rho}\Pi^{\nu\sigma}+
	\Pi^{\mu\sigma}\Pi^{\nu\rho})-\Pi^{\mu\nu}\Pi^{\rho\sigma} \ , \quad 
	\tilde{\mG}^{\mu\nu\rho\sigma}
=\frac{1}{2}(\Sigma^{\mu\rho}\Sigma^{\nu\sigma}+
\Sigma^{\mu\sigma}\Sigma^{\nu\rho})-\Sigma^{\mu\nu}\Sigma^{\rho\sigma} \ . \nonumber \\
\end{eqnarray}
Introducing auxiliary fields $\xi_{\mu\nu}$ and $\psi_{\mu\nu}$ we can rewrite the action (\ref{PULmetric}) in an equivalent form 
\begin{eqnarray}\label{PULmetricaux}
&&	S=\frac{c^4}{16\pi G_g} \int d^{d+1}x \mE 
	[-\frac{c^2}{4}\xi_{\mu\nu}\mG^{\mu\nu\rho\sigma}\xi_{\rho\sigma}+\xi_{\mu\nu}\mG^{\mu\nu\rho\sigma}\mK_{\rho\sigma}
	+\Pi^{\lambda \nu}\stackrel{(C)}{R}_{\lambda\nu}
	+c^2
	\frac{1}{4}B^g_{\mu\nu}B^{\mu\nu}_g]+\nonumber \\
&&	+ \frac{c^4}{16\pi G_f}
	\int d^{d+1}x \mF 
	[-\frac{c^2}{4}\psi_{\mu\nu}\tilde{\mG}^{\mu\nu\rho\sigma}\psi_{\rho\sigma}+\psi_{\mu\nu}\tilde{\mG}^{\mu\nu\rho\sigma}
	\mM_{\rho\sigma}
	+	\Sigma^{\lambda\nu}\stackrel{(D)}{R}_{\lambda\nu}+c^2
	\frac{1}{4}B^f_{\mu\nu}B^{\mu\nu}_f]	
	-
	\nonumber \\
&&	-\frac{c^4}{64\pi G} \int d^{d+1}x \sum_{n=0}^4 \beta_n \mE
	S_n(M) \ .  \nonumber \\
\end{eqnarray}
The action (\ref{PULmetricaux}) is equivalent to the action 
(\ref{PULmetric}) when we take into account the fact that the
equations of motion for $\xi_{\mu\nu}$ and $\psi_{\mu\nu}$ that follow from (\ref{PULmetricaux}) have the form 
\begin{equation}
	-\frac{c^2}{2}\chi_{\mu\nu}+\mK_{\mu\nu}=0 \ , \quad 
	-\frac{c^2}{2}\psi_{\mu\nu}+\mM_{\mu\nu}=0 \ .
\end{equation}
Solving these equations for $\chi$ and $\psi$ respectively and then
inserting this result
 into (\ref{PULmetricaux}) we easily find that it reduces to the action 
(\ref{PULmetric}). However the form of the action (\ref{PULmetricaux}) is suitable for implementing the limit  
$c\rightarrow 0$ (keeping the overall factor $c^4$) with the result
\begin{eqnarray}
&&	S^{(4)}_{NLO}=\frac{1}{16\pi G_g} \int d^{d+1}x \mE 
[\tilde{\xi}^{\mu\nu}\mK_{\mu\nu}
+\Pi^{\lambda \nu}\stackrel{(C)}{R}_{\lambda\nu}+
\nonumber \\
&&+ \frac{1}{16\pi G_f}
\int d^{d+1}x \mF 
[\tilde{\psi}^{\mu\nu}
\mM_{\mu\nu}
+	\Sigma^{\lambda\nu}\stackrel{(D)}{R}_{\lambda\nu}]	
-
\nonumber \\
&&-\frac{1}{64\pi G} \int d^{d+1}x \sum_{n=0}^4 \beta_n \mE
S_n(M) \ ,  \nonumber \\
\end{eqnarray}
where we defined
\begin{equation}
	\xi_{\mu\nu}\mG^{\mu\nu\rho\sigma}=\tilde{\chi}^{\mu\nu} \ , 
	\quad 
	\psi_{\mu\nu}\tilde{\mG}^{\mu\nu\rho\sigma}\equiv \tilde{\psi}^{\mu\nu} \ . 
\end{equation}
Then we can define magnetic Carrollian limit of bimetric gravity as 
\begin{eqnarray}
&&	\mL_{mag}=\lim_{c\rightarrow 0}S^{(4)}_{NLO}=
\frac{1}{16\pi G_g}  \sqrt{-\det (-\tau_\mu\tau_\nu+h_{\mu\nu})} 
[\tilde{\xi}^{\mu\nu}K_{\mu\nu}
+h^{\mu\nu}\stackrel{(C)}{R}_{\lambda\nu}+
\nonumber \\
&&+ \frac{1}{16\pi G_f}
\sqrt{-\det(-u_\mu u_\nu+k_{\mu\nu})} 
[\tilde{\psi}^{\mu\nu}
M_{\mu\nu}
+	\Sigma^{\lambda\nu}\stackrel{(D)}{R}_{\lambda\nu}]	
-
\nonumber \\
&&-\frac{1}{64\pi G_g} \sum_{n=0}^4 \beta_n \sqrt{-\det (-\tau_\mu\tau_\nu+h_{\mu\nu})}
S_n(m^\mu_{ \ \nu}) \ ,  \nonumber \\
\end{eqnarray}	
and we see that in case of the magnetic case only the interaction 
term between two metric is kept in the action. 
In fact, $\tilde{\chi}^{\mu\nu}$ and $\tilde{\psi}^{\mu\nu}$ are
Lagrange multipliers that ensure that the dynamics of the metric  components is trivial.

 {\bf Acknowledgment:}
\\
This work  is supported by the grant “Dualitites and higher order derivatives” (GA23-06498S) from the Czech Science Foundation (GACR).


\end{document}